\documentstyle[12pt,epsf,rotating]{article}

\catcode`\@=11
\long\def\@makefntext#1{
\protect\noindent \hbox to 3.2pt {\hskip-.9pt
$^{{\ninerm\@thefnmark}}$\hfil}#1\hfill}		%CAN BE USED

\def\@makefnmark{\hbox to 0pt{$^{\@thefnmark}$\hss}}  %ORIGINAL

\def\ps@myheadings{\let\@mkboth\@gobbletwo
\def\@oddhead{\hbox{}
\rightmark\hfil\ninerm\thepage}
\def\@oddfoot{}\def\@evenhead{\ninerm\thepage\hfil
\leftmark\hbox{}}\def\@evenfoot{}
\def\sectionmark##1{}\def\subsectionmark##1{}}

\renewcommand{\thefootnote}{\fnsymbol{footnote}}

\newcounter{sectionc}\newcounter{subsectionc}\newcounter{subsubsectionc}
\renewcommand{\section}[1] {\vspace*{0.6cm}\addtocounter{sectionc}{1}
\setcounter{subsectionc}{0}\setcounter{subsubsectionc}{0}\noindent
	{\normalsize\bf\thesectionc. #1}\par\vspace*{0.4cm}}
\renewcommand{\subsection}[1] {\vspace*{0.6cm}\addtocounter{subsectionc}{1}
	\setcounter{subsubsectionc}{0}\noindent
	{\normalsize\it\thesectionc.\thesubsectionc. #1}\par\vspace*{0.4cm}}
\renewcommand{\subsubsection}[1]
{\vspace*{0.6cm}\addtocounter{subsubsectionc}{1}
	\noindent {\normalsize\rm\thesectionc.\thesubsectionc.\thesubsubsectionc.
	#1}\par\vspace*{0.4cm}}

\newcounter{appendixc}
\newcounter{subappendixc}[appendixc]
\newcounter{subsubappendixc}[subappendixc]

\renewcommand{\appendix}[1] {\vspace*{0.6cm}
        \refstepcounter{appendixc}
        \setcounter{figure}{0}
        \setcounter{table}{0}
        \setcounter{equation}{0}
        \renewcommand{\thefigure}{\Alph{appendixc}.\arabic{figure}}
        \renewcommand{\thetable}{\Alph{appendixc}.\arabic{table}}
        \renewcommand{\theappendixc}{\Alph{appendixc}}
        \renewcommand{\theequation}{\Alph{appendixc}.\arabic{equation}}
        \noindent{\bf Appendix \theappendixc #1}\par\vspace*{0.4cm}}

\def\abstracts#1{{

\centering{\begin{minipage}{12.2truecm}\footnotesize\baselineskip=12pt\noindent
	\centerline{\footnotesize ABSTRACT}\vspace*{0.3cm}
	\parindent=0pt #1
	\end{minipage}}\par}}

\renewenvironment{thebibliography}[1]
	{\begin{list}{\arabic{enumi}.}
	{\usecounter{enumi}\setlength{\parsep}{0pt}
\setlength{\leftmargin 1.25cm}{\rightmargin 0pt}
	 \setlength{\itemsep}{0pt} \settowidth
	{\labelwidth}{#1.}\sloppy}}{\end{list}}

\newcounter{itemlistc}
\newcounter{romanlistc}
\newcounter{alphlistc}
\newcounter{arabiclistc}

\newcommand{\fcaption}[1]{
        \refstepcounter{figure}
        \setbox\@tempboxa = \hbox{\footnotesize Fig.~\thefigure. #1}
        \ifdim \wd\@tempboxa > 6in
           {\begin{center}
        \parbox{6in}{\footnotesize\baselineskip=12pt Fig.~\thefigure. #1}
            \end{center}}
        \else
             {\begin{center}
             {\footnotesize Fig.~\thefigure. #1}
              \end{center}}
        \fi}

\newcommand{\tcaption}[1]{
        \refstepcounter{table}
        \setbox\@tempboxa = \hbox{\footnotesize Table~\thetable. #1}
        \ifdim \wd\@tempboxa > 6in
           {\begin{center}
        \parbox{6in}{\footnotesize\baselineskip=12pt Table~\thetable. #1}
            \end{center}}
        \else
             {\begin{center}
             {\footnotesize Table~\thetable. #1}
              \end{center}}
        \fi}

\def\@citex[#1]#2{\if@filesw\immediate\write\@auxout
	{\string\citation{#2}}\fi
\def\@citea{}\@cite{\@for\@citeb:=#2\do
	{\@citea\def\@citea{,}\@ifundefined
	{b@\@citeb}{{\bf ?}\@warning
	{Citation `\@citeb' on page \thepage \space undefined}}
	{\csname b@\@citeb\endcsname}}}{#1}}

\newif\if@cghi
\def\cite{\@cghitrue\@ifnextchar [{\@tempswatrue
	\@citex}{\@tempswafalse\@citex[]}}
\def\citelow{\@cghifalse\@ifnextchar [{\@tempswatrue
	\@citex}{\@tempswafalse\@citex[]}}
\def\@cite#1#2{{$\null^{#1}$\if@tempswa\typeout
	{IJCGA warning: optional citation argument
	ignored: `#2'} \fi}}

 1
 1
 1

\font\ninerm=cmr9

\catcode`@=12

\newcommand{\lsim}{\raisebox{-0.13cm}{~\shortstack{$<$ \\[-0.07cm] $\sim$}}~}

\begin{document}

\renewcommand{\topfraction}{0.8}
\renewcommand{\bottomfraction}{0.8}

\vskip-2cm
\begin{flushright}
DESY 95--073 \\
hep-ph/9504339 \\
April 1995 \\
\end{flushright}
\vskip1cm

\centerline{\normalsize\bf QCD CORRECTIONS TO HIGGS BOSON
DECAYS\footnote{Invited talk presented at the Ringberg Workshop
``Perspectives for Electroweak Interactions in $e^+e^-$ Collisions'',
February 5 - 8, 1995}}
\baselineskip=22pt
\centerline{\footnotesize MICHAEL SPIRA}
\baselineskip=13pt
\centerline{\footnotesize\it II.Institut f\"ur Theoretische
Physik\footnote[3]{Supported by Bundesministerium
f\"ur Bildung und Forschung (BMBF), Bonn, Germany, under Contract 05
6 HH 93P (5) and by EEC Program {\it Human Capital and Mobility} through
Network {\it Physics at High Energy Colliders} under Contract CHRX--CT93--0357
(DG12 COMA).}, Universit\"at Hamburg, Luruper Chaussee 149}
\baselineskip=12pt
\centerline{\footnotesize\it 22761 Hamburg, Germany}
\centerline{\footnotesize E-mail: spira@desy.de}

\vspace*{0.9cm}
\abstracts{
The two--loop ${\cal O}(\alpha_s G_F m_t^2)$ corrections to the $b\bar b$
decay rate of the Standard Model Higgs boson as well as its production via
$e^+e^- \to ZH$ will be presented. These QCD corrections are obtained by using
a low-energy theorem for light Higgs bosons compared to the top quark mass.
The results yield strong
screening effects of the ${\cal O}(G_F m_t^2)$ contributions. After that the
two-loop QCD corrections to the $\gamma\gamma$ and gluonic decays of the
Higgs bosons of the Standard Model and its minimal supersymmetric extension
are discussed. While the corrections to the $\gamma\gamma$ decays remain small
of ${\cal O}(\alpha_s)$ they are huge $\sim$ 50 -- 70 \% in the case of the
gluonic decays.}

\normalsize\baselineskip=15pt
\setcounter{footnote}{0}
\renewcommand{\thefootnote}{\alph{footnote}}
\section{Introduction}
%        ============
\subsection{Standard Model [${\cal SM}$]}
%           ============================
The ${\cal SM}$ contains one Higgs doublet leading to the existence of
one elementary scalar [${\cal CP}$-even] Higgs boson $H$ after absorbing the
three would-be-Goldstone bosons by the $W$ and $Z$ bosons due to the Higgs
mechanism of spontaneous symmetry breaking \cite{higgs}. The only unknown
parameter in the ${\cal SM}$ is the Higgs mass. The failure of experiments at
LEP1 and SLC to detect the decay $Z\to Hf\bar f$ rules out the Higgs mass range
$m_H \lsim 64.3$ Ge$\!$V \cite{janot}. Theoretical analyses of the Higgs sector
lead to the consequence that above a cut-off scale $\Lambda$ the
${\cal SM}$ becomes strongly interacting due to the Higgs four point
coupling exceeding any limit. Requiring unitarity for the ${\cal SM}$
one is left with a consistent formulation of the model up to this cut-off
$\Lambda$, which leads to an upper bound on the Higgs mass.
For a minimal cut-off $\Lambda \sim 1$ Te$\!$V this upper
bound amounts to about 800 Ge$\!$V \cite{higbou}, whereas for the ${\cal SM}$
being weakly interacting up to the GUT scale $\Lambda\sim 10^{15}$ Ge$\!$V
this value comes down to about 200 Ge$\!$V. On the other hand, requiring the
${\cal SM}$ vacuum to be stable, places a lower bound on the Higgs mass
depending on the top quark mass $m_t=176 \pm 13$ Ge$\!$V \cite{cdf} and
the cut-off $\Lambda$. For
$\Lambda\sim 1$ Te$\!$V the Higgs mass has to exceed about 55 Ge$\!$V and
for $\Lambda\sim 10^{15}$ Ge$\!$V the value $\sim 130$ Ge$\!$V \cite{vacstab}.
These bounds decrease dramatically, if the ${\cal SM}$ vacuum is required
to be metastable \cite{vacmeta}.

For Higgs masses below about 135 Ge$\!$V the ${\cal SM}$ Higgs boson
predominantly decays into $b\bar b$ pairs. Consequently this decay mode
determines the signature of the Higgs boson in the lower part of the
intermediate mass range $m_W \lsim m_H \lsim 2 m_Z$.

The gluonic decay $H\to gg$ can be detected at future $e^+e^-$
colliders \cite{eehgg}. Its branching
ratio amounts to $\lsim 10^{-1}$ for Higgs masses below $\sim 140$
Ge$\!$V. A fourth generation of heavy quarks would increase this branching
ratio to a size comparable to the $b\bar b$ decay mode. Therefore the
precise knowledge of this decay mode within the minimal ${\cal SM}$ is
mandatory to disentangle novel effects of new physics from the standard
profile of the Higgs particle.

The rare decay mode $H\to\gamma\gamma$ with a branching ratio of about
$10^{-3}$ for Higgs masses $m_H\lsim 150$ Ge$\!$V yields the main signature
for the search of the ${\cal SM}$ Higgs particle at the LHC for masses
below about 130 Ge$\!$V \cite{hgagalhc,kunzwir}. Higgs production via photon
fusion $\gamma\gamma \to H$ is the relevant mechanism at future high
energy photon colliders \cite{gamfus}.

\subsection{Minimal Supersymmetric Extension of the Standard Model
%           ======================================================
 [${\cal MSSM}$]}
%===============
The ${\cal MSSM}$ requires the
introduction of two Higgs doublets leading to the existence of five
elementary Higgs particles after absorbing the three would-be-Goldstone
bosons via the Higgs mechanism of spontaneous symmetry breaking. These
consist of two neutral scalar [${\cal CP}$-even] ones $h,H$, one neutral
pseudoscalar [${\cal CP}$-odd] $A$ and two charged ones $H^\pm$. At tree level
the Higgs sector can be described by two basic parameters that are usually
chosen to be (i) $\mbox{tg}\beta = v_2/v_1$ with $v_1, v_2$ being the
vacuum expectation values of the neutral scalar Higgs states, and (ii)
one of the Higgs masses, usually the pseudoscalar mass $m_A$. After
fixing these two parameters all others are determined due to constraints
required by supersymmetry. One of these sets an upper bound on the
mass of the lightest neutral scalar Higgs boson $h$, which must be lighter
than the $Z$ boson at tree level. This value increases significantly
by the inclusion of radiative corrections, with the leading part
increasing as the fourth power of the top mass $m_t$ \cite{mssmrad},
to about 130 Ge$\!$V.

The Yukawa couplings of the neutral Higgs bosons to the standard fermions
and the couplings to gauge bosons are modified compared to the ${\cal SM}$ by
additional
coefficients fixed by the angle $\beta$ and the mixing angle $\alpha$ of
the neutral scalar Higgs particles $h,H$. These couplings are shown in Table
\ref{tb:coup} relative to the ${\cal SM}$ couplings. An important feature
is the absence of any pseudoscalar $A$ coupling to gauge bosons at
tree level.
\begin{table}[hbt]
\tcaption{\label{tb:coup}
{Higgs couplings in the ${\cal MSSM}$ to fermions and gauge bosons
relative to ${\cal SM}$ couplings.}}
\renewcommand{\arraystretch}{1.5}
\begin{center}
\begin{tabular}{|lc||ccc|} \hline
\multicolumn{2}{|c||}{$\phi$} & $t$ & $b$ & $V=W,Z$ \\ \hline \hline
${\cal SM}$ & $H$ & 1 & 1 & 1 \\ \hline
${\cal MSSM}$ & $h$ & $\cos\alpha/\sin\beta$ & $-\sin\alpha/\cos\beta$ &
$\sin(\beta-\alpha)$ \\
& $H$ & $\sin\alpha/\sin\beta$ & $\cos\alpha/\cos\beta$ &
$\cos(\beta-\alpha)$ \\
& $A$ & $ 1/\mbox{tg$\beta$}$ & $\mbox{tg$\beta$}$ & 0 \\ \hline
\end{tabular} \\[0.3cm]
\renewcommand{\arraystretch}{1.2}
\end{center}
\end{table}

The direct search for the Higgs particles at LEP1 excludes the mass ranges
$m_{h,H}\lsim 45$ Ge$\!$V for the neutral scalar, $m_A\lsim 25$ Ge$\!$V for
the neutral pseudoscalar and $m_{H^\pm}\lsim 45$ Ge$\!$V for the charged
Higgs bosons \cite{lepmssm}.

The main decay modes of the neutral Higgs particles are in general $b\bar b$
decays [$\sim 90$ \%] and $\tau^+\tau^-$ decays [$\sim 10$ \%]. The gluonic
decay mode can reach a branching ratio of a few percent for the light scalar
$h$, with a mass close to its upper end, and the pseudoscalar $A$ as well as
the heavy scalar $H$ just below the $t\bar t$ threshold for small tg$\beta$.

Rare $\gamma\gamma$ decays of the neutral scalar Higgs bosons provide the
most important signature in the main part of the ${\cal MSSM}$ parameter
space at the LHC \cite{kunzwir}.

This paper is organized as follows. In Section 2 we discuss the derivation of
the two-loop ${\cal O}(\alpha_s G_F m_t^2)$ corrections to the $H\to b\bar b$
decay mode of the Standard Higgs boson. Section 3 presents the analogous
corrections to Standard Higgs production via $e^+e^-\to ZH$ at LEP. In
Section 4 we describe the calculation of the two-loop QCD corrections to the
photonic decays $\Phi\to \gamma\gamma$ of the Higgs particles in the
${\cal SM}$ and the ${\cal MSSM}$ and in Section 5 the corresponding ones
to the gluonic decays $\Phi\to gg$.

\section{$H\to b\bar b$ [${\cal SM}$]}
%        ============================
To derive the two-loop ${\cal O}(\alpha_s G_F m_t^2)$ corrections to the
$H\to b\bar b$ decay width we take advantage of a low-energy theorem for
light Higgs bosons. This theorem is derived in the limit of vanishing Higgs
momentum, where the Higgs field acts as a constant c-number, because
$[{\cal P}_\mu, H] = i\partial_\mu H = 0$ with ${\cal P}_\mu$ denoting the
four-momentum operator. Hence the kinetic terms of the Higgs boson in the
basic Lagrangian can be neglected in this limit so that the entire interaction
with matter particles is generated by the mass substitution $m \to m (1+H/v)$
for fermions as well as massive gauge bosons \cite{theorem1,theorem2,theorem}.
The parameter $v=246$ Ge$\!$V
denotes the vacuum expectation value. To extend this theorem to higher orders
in perturbation theory all parameters have to be replaced by their bare
quantities, marked by the index 0, leading finally to the low-energy theorem
\cite{theo.h,theo.h2,hzz}:
\begin{equation}
\lim_{p_H\to 0} {\cal M}(XH) = \sum_{i=f,V} \frac{1}{v_0}
\frac{m_i^0\partial}{\partial m_i^0} {\cal M} (X)
\label{eq:let.H}
\end{equation}
${\cal M}(X)$ denotes the matrix element of any particle configuration $X$
and ${\cal M}(XH)$ the corresponding one with an external Higgs particle added.
Using eq.(\ref{eq:let.H}) one can build up effective
Lagrangians describing the matrix elements ${\cal M}(XH)$. One important
subtelty for the application of the theorem is that all mass dependent
couplings $g_i^0 = m_i^0/v_0$, which are generated by the mass substitution,
have
to be kept fixed with respect to mass differentiation, so that only dynamical
masses in the propagators will be affected by the differentiation in
eq.(\ref{eq:let.H}).

For the derivation of the two-loop corrections of ${\cal O}(G_F m_t^2)$ to
the $H\to b\bar b$ decay we have to compute the corresponding corrections
to the $b$ propagator:
\begin{equation}
{\cal M} (b\to b) = m_b^0 \left[ 1+\Sigma_S(0) \right] + \not \! p \left[
\Sigma_V (0) + \gamma_5 \Sigma_A (0) \right]
\label{eq:btob}
\end{equation}
In the calculation the $b$ mass has to be put equal to zero inside the
loops and kept finite only as an overall coefficient. Furthermore we may
neglect
the $W$ mass and take into account the longitudinal components $w^\pm$ only
to compute the ${\cal O}(\alpha_s G_F m_t^2)$ correction. Applying the
low-energy theorem we can derive the effective coupling of the Higgs boson $H$
to $b$ quarks at the same order:
\begin{equation}
\lim_{p_H\to 0} {\cal M} (b\to bH) = \frac{1}{v_0} \left( \frac{m_b^0\partial}
{\partial m_b^0} + \frac{m_t^0\partial}{\partial m_t^0} \right) {\cal M}
(b\to b)
\end{equation}
The calculation of the two-loop diagrams yields the following results
for the different pieces of the $b$ self-energy in $n=4-2\epsilon$ dimensions
\cite{theo.h2}:
\begin{eqnarray}
m_b^0 \Sigma_S (0) & = & g_b^0 g_t^0 m_t^0 \frac{\Gamma(1+\epsilon)}{(4\pi)^2}
\left( \frac{4\pi \mu^2}{(m_t^0)^2}\right)^\epsilon \left\{ \frac{2}{\epsilon}
+ 2 + 2\epsilon \right\} \nonumber \\
& + & C_F \frac{\alpha_s}{\pi} g_t^0
\frac{\Gamma^2(1+\epsilon)}{(4\pi)^2} \left( \frac{4\pi
\mu^2}{(m_t^0)^2}\right)^{2\epsilon} \left\{ \frac{3}{2\epsilon^2}g_b^0 m_t^0
\right. \nonumber \\
& & \hspace{4.5cm} \left. + \frac{1}{\epsilon} \left[ 2g_b^0m_t^0 -
\frac{3}{4} g_t^0 m_b^0 \right] + {\cal O} (1) \right\} \nonumber \\
\Sigma_V (0) & = & (g_t^0)^2 \frac{\Gamma(1+\epsilon)}{(4\pi)^2}
\left( \frac{4\pi \mu^2}{(m_t^0)^2}\right)^\epsilon \left\{
-\frac{1}{2\epsilon} - \frac{3}{4} - \frac{7}{8}\epsilon \right\} \nonumber \\
& + & C_F \frac{\alpha_s}{\pi} (g_t^0)^2
\frac{\Gamma^2(1+\epsilon)}{(4\pi)^2} \left( \frac{4\pi
\mu^2}{(m_t^0)^2}\right)^{2\epsilon} \left\{ -\frac{3}{8\epsilon^2} -
\frac{1}{8\epsilon} + {\cal O} (1) \right\}
\label{eq:bself1}
\end{eqnarray}
with $g_q^0 = m_q^0/v_0~(q=t,b)$ and $C_F = 4/3$ denoting the corresponding
Yukawa couplings and color factor. After taking the derivative with respect
to the top
and bottom masses we have to perform the renormalization of the bare couplings,
wave functions and masses. For this purpose we have adopted the on-shell
renormalization scheme, which fixes the counter terms as:
\begin{eqnarray}
m_b^0 & = & m_b [1-\Sigma_S(0) - \Sigma_V (0)] \nonumber \\
b_0   & = & [1 + \frac{1}{2} \Sigma_V (0)] b \nonumber \\
m_t^0 & = & m_t \left( 1-\frac{\delta m_t}{m_t} \right) \nonumber \\
\frac{\delta m_t}{m_t} & = & C_F \frac{\alpha_s}{\pi}
\left( \frac{4\pi
\mu^2}{m_t^2}\right)^\epsilon \Gamma(1+\epsilon) \left\{
\frac{3}{4\epsilon} + 1 + 2\epsilon \right\} \nonumber \\
\frac{H_0}{v_0} & = & \frac{H}{v} (1+\delta_u)
\end{eqnarray}
with the universal correction \cite{delta.u,hbb.2}
\begin{equation}
\delta_u = x_t \left\{ \frac{7}{2} - \frac{3}{4} \left[ 3+2\zeta(2)\right]
C_F \frac{\alpha_s}{\pi} \right\}
\end{equation}
where $x_t = G_F m_t^2 / (8\sqrt{2}\pi^2)$. Finally we end up with the
effective Lagrangian \cite{theo.h2},
\begin{eqnarray}
{\cal L}_{eff} & = & -m_b \bar b b \frac{H}{v} [ 1+\delta_{nu}][ 1+\delta_u ]
\nonumber \\
\delta_{nu} & = & x_t \left( -3 + \frac{3}{4} C_F \frac{\alpha_s}{\pi} \right)
\end{eqnarray}
which has to be considered as the basic Lagrangian of the modified theory with
the heavy top quark being integrated out so that the perturbative corrections
due to the interaction among the light particles have to be added to gain
the full correction to the $H\to b\bar b$ process. These corrections coincide
with the well-known one-loop QCD corrections \cite{hbb.qcd}, so that the
total correction to the decay width reads \cite{theo.h2,hbb.2}:
\begin{eqnarray}
\Gamma (H\to b\bar b) & = & \Gamma_{LO} (H\to b\bar b) (1+\delta)
(1+\delta_{QCD}) \nonumber \\
\Gamma_{LO} (H\to b\bar b) & = & \frac{N_c G_F m_H m_b^2}{4\sqrt{2}\pi}
\sqrt{1-4\frac{m_b^2}{m_H^2}} \nonumber \\
\delta & = & x_t \left( 1 - 3 \left[ 1 + \zeta(2) \right] C_F
\frac{\alpha_s}{\pi} \right) \nonumber \\
\delta_{QCD} & \longrightarrow & C_F \frac{\alpha_s}{\pi} \left\{ \frac{9}{4}
- \frac{3}{2} \log \frac{m_H^2}{m_b^2} \right\} \hspace{1cm} (m_b \ll m_H)
\end{eqnarray}
The large logarithm of the $\delta_{QCD}$ part can be absorbed into the
running $b$ mass of the lowest order decay width by changing the scale
from the $b$ mass itself to the Higgs mass $m_H$ \cite{hbb.qcd}. The
correction $\delta$ is numerically given by
\begin{equation}
\delta \approx x_t \left\{ 1 - 3.368 \alpha_s \right\}
\end{equation}
yielding a screening effect of about 40\% in the leading top mass
term, in agreement with the general observation of screening in all known
${\cal O}(\alpha_s G_F m_t^2)$ corrections to physical observables in the
${\cal SM}$.

\section{$e^+e^-\to ZH$ [${\cal SM}$]}
%        ============================
To obtain the two-loop ${\cal O}(\alpha_s G_F m_t^2)$ corrections to Higgs
production via $e^+e^-\to ZH$ the low-energy theorem eq.(\ref{eq:let.H})
can again be used. The effective coupling of a light Higgs boson with a mass
negligible as compared to the top quark mass $m_t$ can be derived from the
corresponding corrections to the on-shell $Z$ self-energy:
\begin{equation}
{\cal M}(Z\to Z) = (M^0_Z)^2 + (g_Z^0 v_0)^2 \Pi_{ZZ}(0)
\label{eq:MZZ}
\end{equation}
Calculating the two-loop corrections to the self-energy $\Pi_{ZZ}$ for large
top masses at vanishing $Z$ mass and $Z$ momentum one arrives at the result
\cite{drho2,pizz}
\begin{eqnarray}
\Pi_{ZZ} (0) & = & x_t^0 \left( \frac{4\pi\mu^2}{(m_t^0)^2} \right)^\epsilon
\Gamma(1+\epsilon) \frac{6}{\epsilon} \nonumber \\
& + & C_F \frac{\alpha_s}{\pi} x^0_t \left(
\frac{4\pi\mu^2}{(m_t^0)^2} \right)^{2\epsilon} \Gamma^2(1+\epsilon) \left\{
\frac{9}{2\epsilon^2} - \frac{21}{4\epsilon} + \frac{3}{8} \right\}
\end{eqnarray}
with
\begin{equation}
x_t^0 = \frac{G_F^0 (m_t^0)^2}{8\sqrt{2}\pi^2} \hspace{3cm} g_Z^0 =
\frac{M_{Z0}}{v_0}
\end{equation}
The low-energy theorem leads to the following relation to the effective
coupling of the Higgs bosons to $Z$ bosons at vanishing Higgs momentum
\begin{equation}
\lim_{p_H\to 0} {\cal M} (Z\to ZH) = \frac{1}{v_0} \left( \frac{M_Z^0\partial}
{\partial M_Z^0} + \frac{m_t^0\partial}{\partial m_t^0} \right) {\cal M}
(Z\to Z)
\label{eq:MHZZ}
\end{equation}
After taking the differentiation and performing the renormalization of the
bare parameters in the on-shell renormalization scheme we end up with the
effective Lagrangian
\begin{eqnarray}
{\cal L}_{HZZ} & = & \left( \sqrt{2}G_F\right)^{1/2} M_Z^2 Z^\mu Z_\mu H
(1+\delta_{HZZ}) \nonumber \\
\delta_{HZZ} & = & x_t \left\{ -\frac{5}{2} + \frac{3}{2}
\left[ \frac{15}{2} - \zeta (2) \right] C_F \frac{\alpha_s}{\pi} \right\}
\label{eq:LHZZ}
\end{eqnarray}
This Lagrangian contains only the ${\cal O}(G_F m_t^2)$ and ${\cal O}
(\alpha_s G_F m_t^2)$ corrections to this coupling.
All other higher-order corrections are omitted from eq.(\ref{eq:LHZZ}).
The correction $\delta_{HZZ}$ amounts to $\delta_{HZZ} = -5x_t/2 (1-4.684
\alpha_s/\pi)$ yielding a screening effect of about 20\% in the leading
top mass contribution to the effective $HZZ$ coupling. In order to derive
the correction to the cross section $\sigma (e^+e^-\to ZH)$ an additional
term due to the renormalization of the $Ze^+e^-$ vertex in the on-shell
scheme has to be added,
\begin{eqnarray}
\sigma (e^+ e^- \to ZH) & = & \sigma_{LO} (e^+ e^- \to ZH) (1+
\delta_{HZe^+e^-}) \nonumber \\
\delta_{HZe^+ e^-} & = & 2\delta_{HZZ} + \left( 1-8 \frac{c_w^2 Q_e v_e}{v_e^2
+ a_e^2} \right) \Delta \rho
\end{eqnarray}
which is proportional to the correction of the $\rho$
parameter \cite{drho2,pizz,drho1}:
\begin{equation}
\Delta \rho = x_t \left\{ 3 - \frac{3}{2} \left[ 1 + 2 \zeta (2) \right]
C_F \frac{\alpha_s}{\pi} \right\}
\end{equation}
The final correction can be cast into the form \cite{hzz}
\begin{equation}
\delta = -2 x_t \left\{ 1
- \left[ \frac{21}{2} - 3 \zeta (2) \right] C_F \frac{\alpha_s}{\pi}
+ 12 \frac{c_w^2 Q_e v_e}{v_e^2 + a_e^2} \left[ 1
- \left[ \frac{1}{2} + \zeta (2) \right] C_F \frac{\alpha_s}{\pi} \right]
\right\}
\end{equation}
Numerically this amounts to a screening effect of about 20\% in the leading
top mass term ${\cal O}(G_F m_t^2)$.

\section{$\Phi\to\gamma\gamma$ [${\cal SM}$, ${\cal MSSM}$]}
%        ==================================================
The lowest order $\Phi\gamma\gamma$ coupling [$\Phi$ denotes all possible
kinds of Higgs bosons within the ${\cal SM}$ and the ${\cal MSSM}$] is
mediated by fermion and $W$
boson loops yielding the following expression for the lowest order decay
width \cite{theorem1,theorem2,agaga}
\begin{equation}
\Gamma(\Phi\rightarrow\gamma\gamma)=
               \frac{G_F\alpha^2m_\Phi^3}{128\sqrt{2}\pi^3}
\left| \sum_f N_c e_f^2 g_f^\Phi A_f(\tau_f) + g_W^\Phi A_W(\tau_W)
\right|^2
\end{equation}
with $g_i^\Phi$ denoting the corresponding couplings of Table \ref{tb:coup}.
The individual amplitudes are given by
\begin{eqnarray}
A_f^H(\tau) & = & 2 \tau [1 +(1-\tau)f(\tau)] \nonumber \\
A_f^A(\tau) & = & 2 \tau f(\tau) \nonumber \\
A_W^H(\tau) & = & - [2 +3\tau+3\tau(2-\tau)f(\tau)]
\end{eqnarray}
where the scaling variables are defined as $\tau_i=4 m^2_i/m_\Phi^2i~(i=f,W)$;
the function $f(\tau)$ can be expressed as
\begin{eqnarray}
f(\tau)=\left\{
\begin{array}{ll}  \displaystyle
\arcsin^2\frac{1}{\sqrt{\tau}} & \tau\geq 1 \\
\displaystyle -\frac{1}{4}\left[ \log\frac{1+\sqrt{1-\tau}}
{1-\sqrt{1-\tau}}-i\pi \right]^2 \hspace{0.5cm} & \tau<1
\end{array} \right.
\label{eq:ftau}
\end{eqnarray}
Heavy particles provide the dominant contributions to this rare decay mode,
so that we restrict ourselves to the $W$, top and bottom contributions in the
following. The branching ratio amounts to about $10^{-3}$ in the mass ranges,
where this decay process is visible. The cross section for Higgs
boson production via photon fusion, the relevant production mechanism at
future high energy photon colliders, can be derived from the decay width
via
\begin{equation}
\sigma(\gamma\gamma\to\Phi) = \frac{8\pi^2}{m_\Phi^3} \Gamma(\Phi\to
\gamma\gamma) \delta \left(1-\frac{m_\Phi^2}{s} \right)
\end{equation}

The two-loop QCD corrections to the photonic decay mode can be parameterized
as a correction to the quark amplitude
\begin{equation}
A_Q = A_Q^{LO} \left[ 1+ C_\Phi (\tau_Q) \frac{\alpha_s}{\pi} \right]
\end{equation}
To evaluate the coefficient $C_\Phi (\tau_Q)$ we reduced the five-dimensional
Feynman integrals of the virtual corrections analytically down to
one-dimensional ones containing trilogarithms in the integrand. The
regularization of ultraviolet singularities is performed in $n=4-2\epsilon$
dimensions. The pseudoscalar $\gamma_5$ coupling is defined in the scheme by
't Hooft and Veltman, which has been systematized by Breitenlohner and Maison
\cite{gamma5}. This definition of $\gamma_5$ in $n$ dimensions reproduces
the axial vector anomaly and is consistent up to any order in perturbation
theory. The renormalization is performed in the on-shell scheme with the
running quark mass defined by the boundary condition
$m_Q^{on} (\mu^2=m_Q^2) = m_Q$,
where $m_Q$ denotes the physical mass defined as the pole of the quark
propagator. This mass definition does {\it not} coincide with the usually
chosen running $\overline{MS}$ mass, but differs by a finite amount
$m_Q^{on}(\mu^2) = m_Q^{\overline{MS}}(\mu^2) \left[ 1+ 4/3~
\alpha_s(m_Q^2)/\pi + {\cal O} (\alpha_s^2(m_Q^2)) \right]$.
In the limit of large quark masses $m_Q$ compared to the Higgs mass $m_\Phi$
the coefficients $C_\Phi$ approach the following values
\cite{theo.h,hgaga.ho,hgaga.ho2}
\begin{equation}
C_H \to -1 \hspace{4cm} C_A \to 0
\end{equation}
These limits can also be derived by using low-energy theorems:
\paragraph{Scalar Higgs bosons.}
%          ====================
To derive the QCD correction to the $H\gamma\gamma$ coupling in the limit
of small Higgs masses we have to differentiate the vacuum polarization function
$\Pi$ by the heavy quark mass $m_Q$. The heavy quark part of this function
can be expressed in terms of the effective Lagrangian
\begin{equation}
{\cal L}_{eff} = -\frac{1}{4} F_0^{\mu\nu} F_{0\mu\nu} \left\{ 1 + \Pi_Q
\left(\frac{\mu^2}{m_Q^2} \right) \right\}
\end{equation}
where $m_Q$ denotes the physical renormalized heavy quark mass. Rewriting the
differentiation by the bare mass $m_Q^0$ in eq.(\ref{eq:let.H}) in terms of
the renormalized mass $m_Q$ a correction due to the anomalous mass dimension
$\gamma_m$ is obtained.
The differentiation of the vacuum polarization function by the renormalized
heavy quark mass yields the heavy quark contribution to the QED $\beta$
function,
\begin{equation}
\frac{m_Q^0\partial}{\partial m_Q^0} \Pi_Q \left( \frac{\mu^2}{m_Q^2} \right)
= \frac{1}{1+\gamma_m} \frac{m_Q\partial}{\partial m_Q} \Pi_Q \left(
\frac{\mu^2}{m_Q^2} \right)
= -\frac{\beta_\alpha^Q/\alpha}{1+\gamma_m}
\end{equation}
so that finally we end up with the effective Lagrangian for the $H\gamma\gamma$
coupling in the limit of a heavy quark $Q$ compared to the Higgs mass
\cite{theo.h,hgg.qcd,hgg.qcd1}
\begin{equation}
{\cal L}_{H\gamma\gamma} = \frac{1}{4} \frac{\beta_\alpha^Q/\alpha}{1+\gamma_m}
F^{\mu\nu} F_{\mu\nu} \frac{H}{v}
\end{equation}
Expanding the $\beta$ function and the anomalous mass dimension up to
${\cal O}(\alpha_s)$
\begin{equation}
\frac{\beta_\alpha^Q}{\alpha} = 2 e_Q^2 \frac{\alpha}{\pi} \left[ 1 +
\frac{\alpha_s}{\pi} \right] \hspace{1cm} \mbox{and} \hspace{1cm}
\gamma_m = 2\frac{\alpha_s}{\pi}
\end{equation}
we arrive at the correction in the limit of light Higgs masses,
\begin{eqnarray}
\frac{m_H^2}{4m_Q^2} \to 0\,: \hspace{1cm} 1+
C_H\frac{\alpha_s}{\pi} \to \frac{1+\alpha_s/\pi} {1+2 \alpha_s/\pi}
\ = 1 - \frac{\alpha_s}{\pi}
\end{eqnarray}
in agreement with the explicit expansion of the two-loop diagrams.

\paragraph{Pseudoscalar Higgs bosons.}
%          ==========================
Also for pseudoscalar Higgs bosons a low-energy theorem can be derived based
on the ABJ-anomaly of the axial vector current $j_5^\mu = \bar Q \gamma^\mu
\gamma_5 Q$ \cite{abj}
\begin{equation}
\partial_\mu j_\mu^5 = 2m_Q \overline{Q} i\gamma_5 Q +
N_c e_Q^2\frac{\alpha}{4\pi} F_{\mu\nu} \widetilde{F}_{\mu\nu}
\label{eq:anomaly}
\end{equation}
with $\widetilde{F}_{\mu\nu} = \epsilon_{\mu\nu\alpha\beta} F_{\alpha\beta}$
denoting the dual field strength tensor. A general theorem states that there
are {\it no} radiative corrections modifying eq.(\ref{eq:anomaly}), which
therefore remains valid up to all orders in perturbation theory \cite{adlbar}.
The operator $\partial_\mu j^\mu_5$, multiplied with
the pseudoscalar field operator $A$ at vanishing momentum, fulfills the
low-energy condition \cite{suther}
\begin{equation}
\lim_{p_A\to 0} \langle \gamma \gamma | \partial_\mu j^\mu_5 A | A
\rangle = 0
\end{equation}
Using the basic interaction Lagrangian ${\cal L}_{int} = -m_Q \bar Q i \gamma_5
Q A/v$ for the coupling of the pseudoscalar $A$ to quarks one immediately
arrives at the effective Lagrangian
\cite{theo.h,hgaga.ho}
\begin{equation}
{\cal L}_{eff} (A \gamma\gamma) = N_c e_Q^2 \frac{\alpha}{8\pi}
F_{\mu\nu} \widetilde{F}_{\mu\nu} \frac{A}{v}
\end{equation}
Because of the Adler-Bardeen theorem [the non-renormalization of the
ABJ-anomaly] \cite{adlbar} this Lagrangian is valid up to all orders in
perturbation theory,
so that the QCD corrections are vanishing in the heavy quark limit
\begin{equation}
\frac{m_A^2}{4m_Q^2} \to 0\,: \hspace{1cm} C_A \to 0
\end{equation}
in agreement with the explicit expansion of the two-loop contributions. \\

\begin{figure}[hbt]

\vspace*{-1.4cm}
\hspace*{-1.60cm}
\begin{turn}{90}%
\epsfxsize=7.5cm \epsfbox{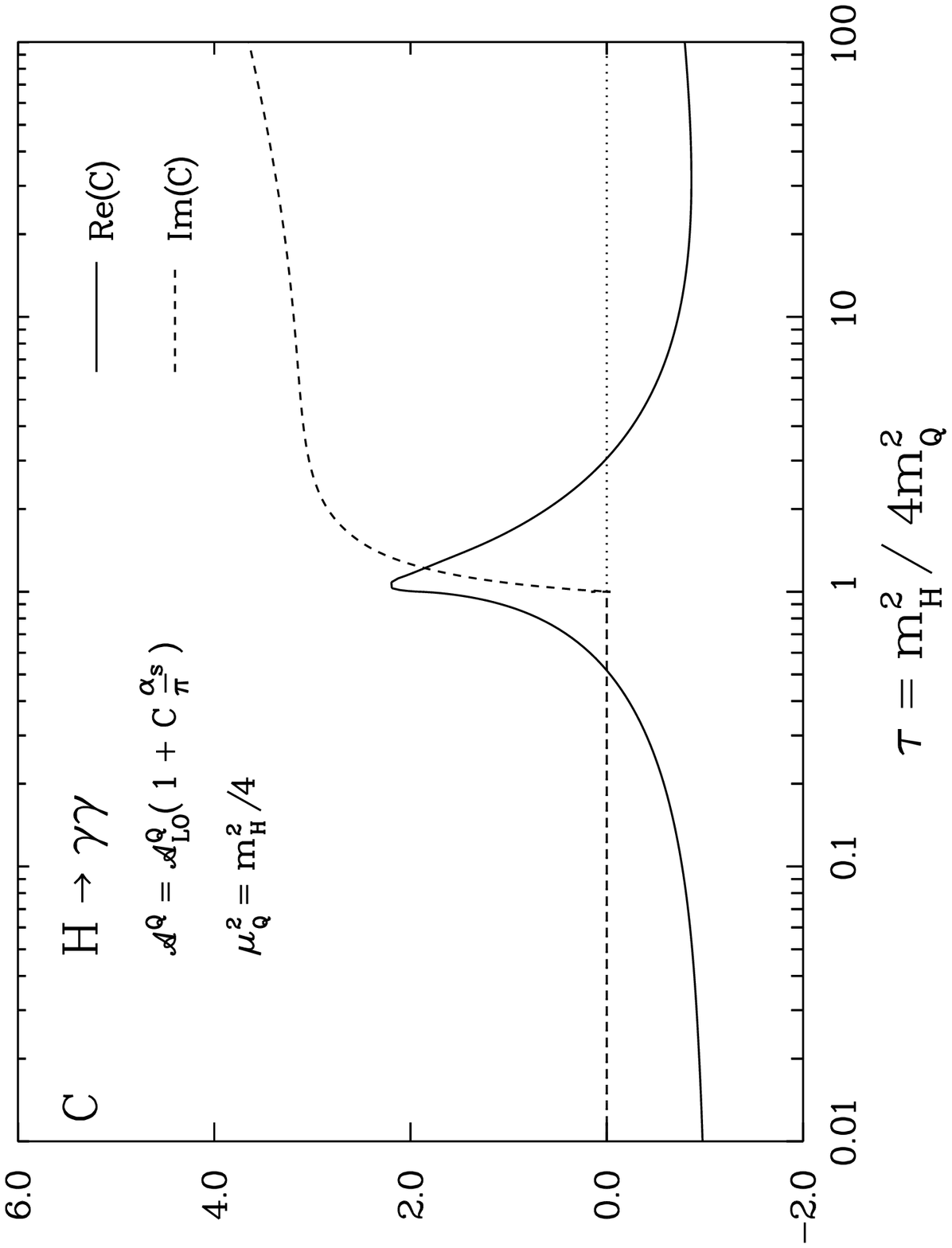}
\end{turn}
\vspace*{-0.8cm}

\vspace*{-6.74cm}
\hspace*{6.40cm}
\begin{turn}{90}%
\epsfxsize=7.5cm \epsfbox{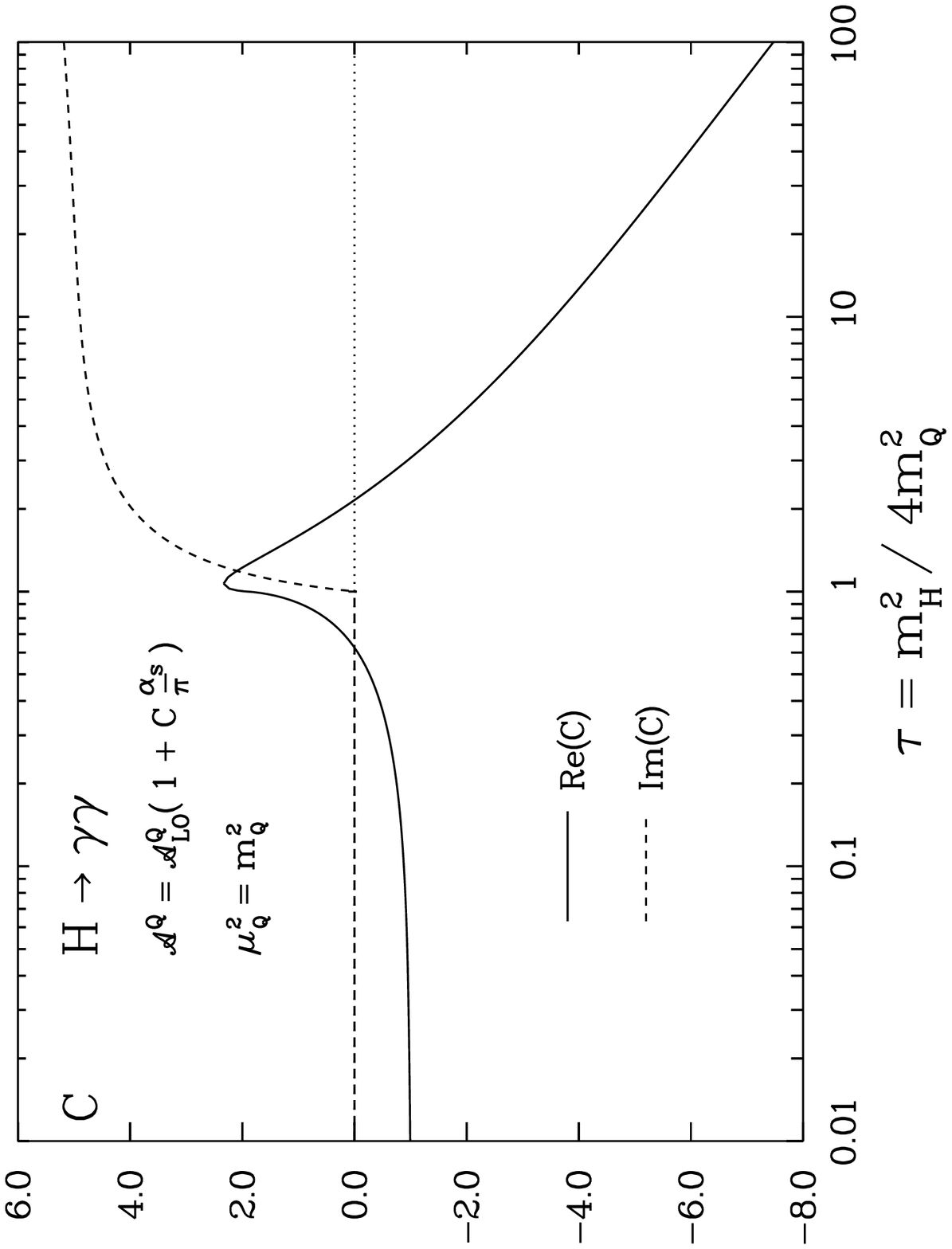}
\end{turn}
\vspace*{-0.4cm}

\fcaption{\label{fg:C.H} Real and imaginary parts of the QCD correction
to the scalar two-photon decay amplitude for two different scales $\mu_Q$ of
the running quark mass.}
\vspace*{-0.2cm}

\end{figure}
\begin{figure}[hbt]

\vspace*{-1.95cm}
\hspace*{0.50cm}
\begin{turn}{90}%
\epsfxsize=10.0cm \epsfbox{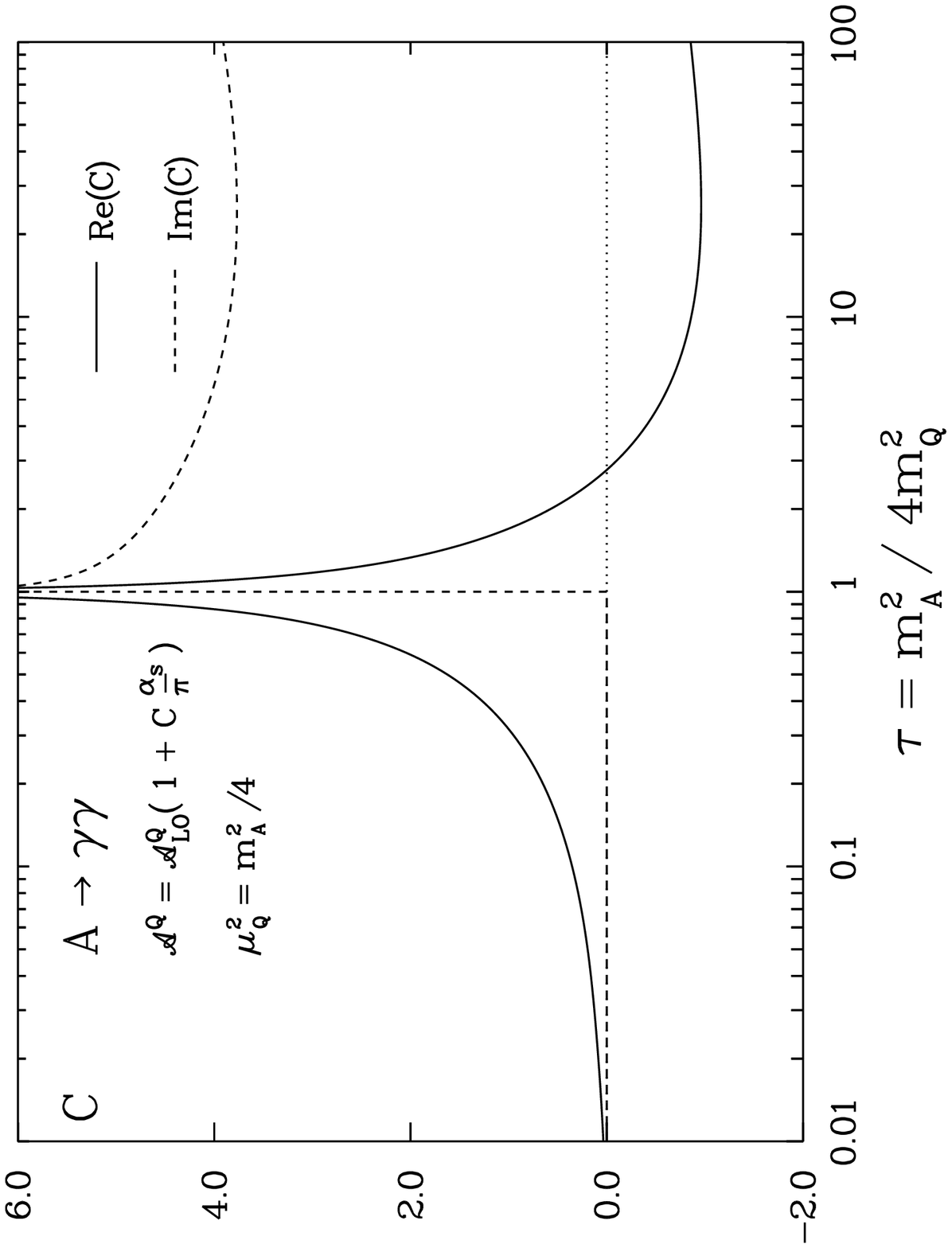}
\end{turn}
\vspace*{-0.6cm}

\fcaption{\label{fg:C.A} Real and imaginary parts of the QCD correction
to the pseudoscalar two-photon decay amplitude.}
\vspace*{-0.5cm}

\end{figure}
In Fig.\ref{fg:C.H} the coefficient $C_H$ is shown as a function of the
scaling variable $\tau=\tau_Q^{-1}$. For large Higgs masses the large
logarithms
can be absorbed into the running quark mass by shifting the scale from the
quark mass itself to the Higgs mass $m_\Phi/2$\footnote{The scale $\mu =
m_\Phi/2$ has been chosen to define the threshold to be at the correct location
$2m_Q(m_Q)$.}. The QCD corrections are small of ${\cal O}(\alpha_s)$ so that
the processes are theoretically under control. The QCD correction to the
pseudoscalar decay develops a Coulomb singularity at threshold [$m_A = 2 m_Q$],
which is due to the equality of the quantum numbers of the pseudoscalar
Higgs bosons and the ground state of heavy quarkonium $(\bar Q Q)$. This
property leads to a step in the imaginary part of $C_A$ and the
corresponding logarithmic singularity in the real part [see
Fig.\ref{fg:C.A}]. Hence the perturbative analysis is not valid
within a margin of a few Ge$\!$V around the threshold, requiring the analysis
to be improved in the threshold region \cite{thresh}.

\begin{figure}[hbt]

\vspace*{-1.95cm}
\hspace*{0.50cm}
\begin{turn}{90}%
\epsfxsize=10.0cm \epsfbox{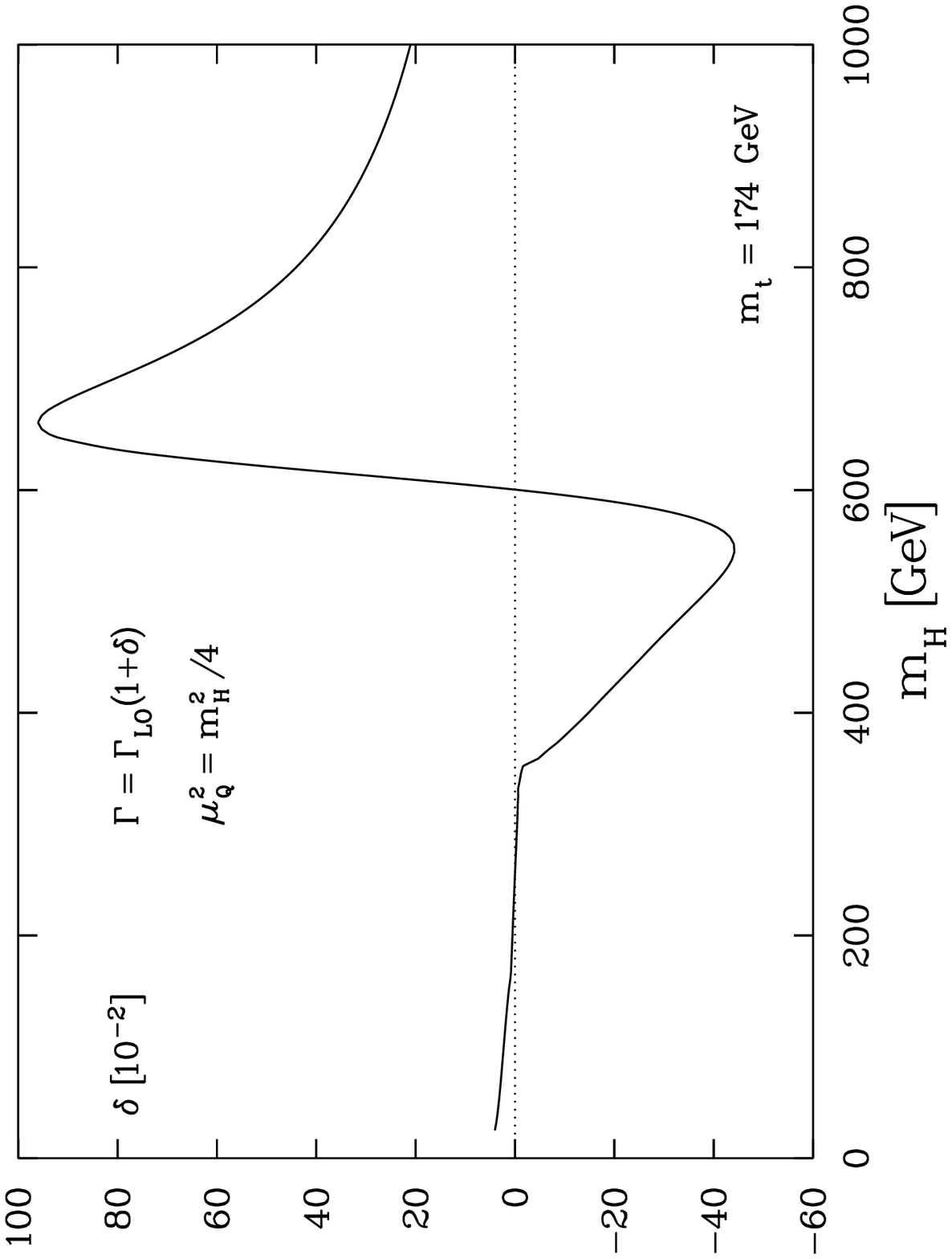}
\end{turn}
\vspace*{-0.6cm}

\fcaption{\label{fg:d.sm} The relative QCD corrections to the photonic decay
width of the ${\cal SM}$ Higgs boson.}
\vspace*{-0.2cm}

\end{figure}
\begin{figure}[hbt]

\vspace*{-1.15cm}
\hspace*{-1.60cm}
\begin{turn}{90}%
\epsfxsize=7.5cm \epsfbox{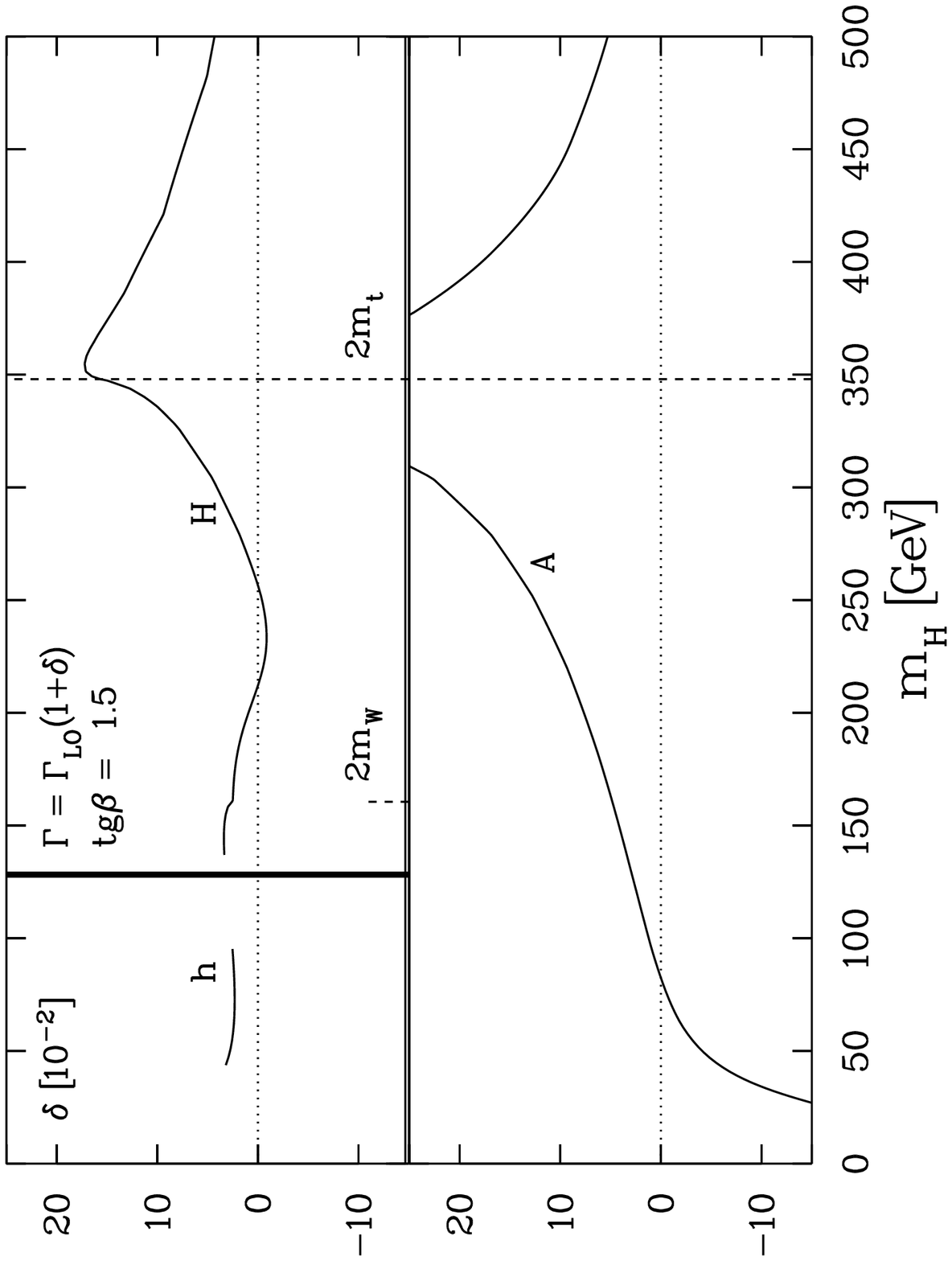}
\end{turn}
\vspace*{-0.8cm}

\vspace*{-6.73cm}
\hspace*{6.40cm}
\begin{turn}{90}%
\epsfxsize=7.5cm \epsfbox{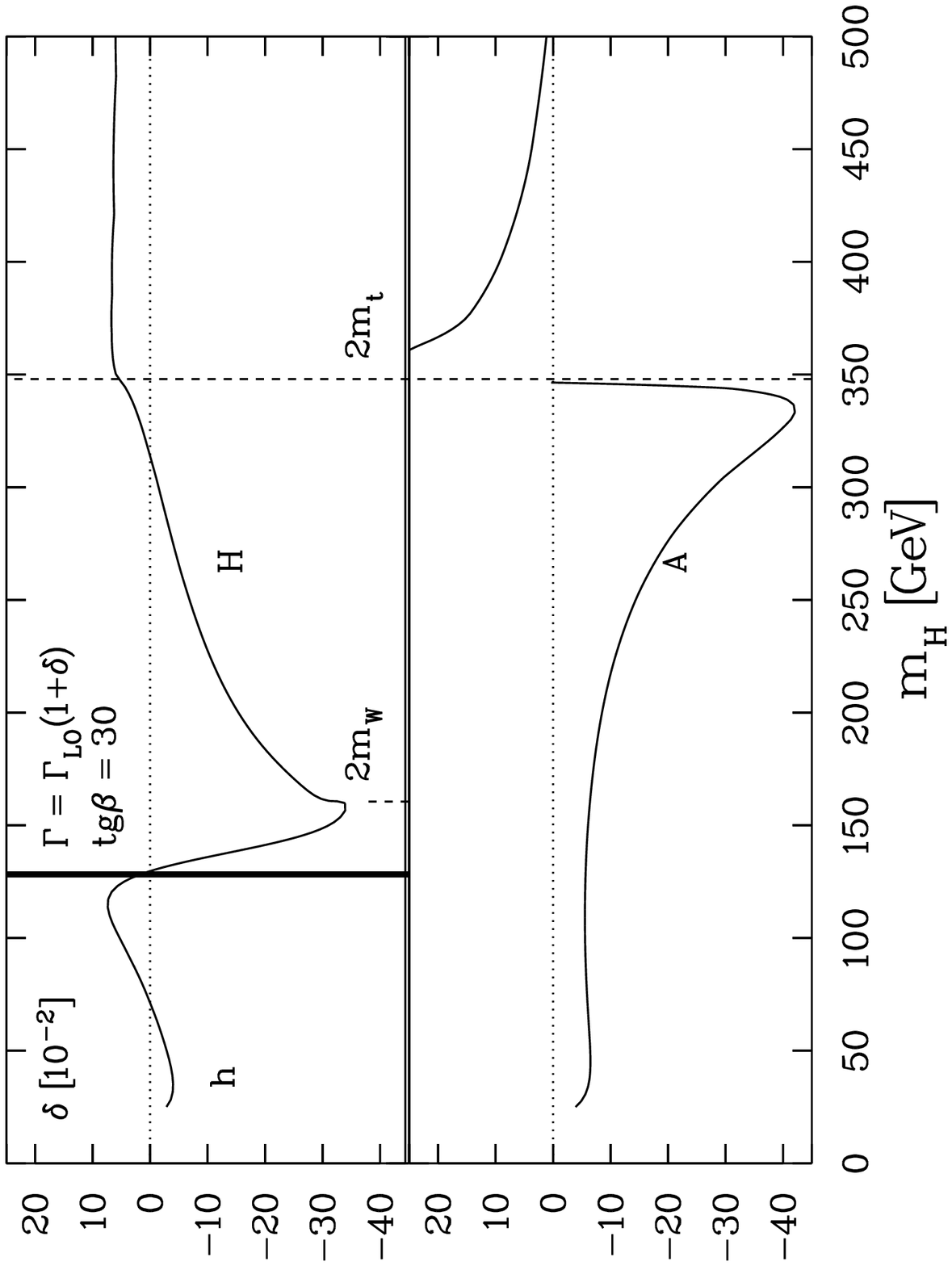}
\end{turn}
\vspace*{-0.7cm}

\fcaption{\label{fg:d.mssm} The relative QCD corrections to the photonic decay
widths of the ${\cal MSSM}$ Higgs bosons for two different values of
tg$\beta$.}
\vspace*{-0.5cm}

\end{figure}
The relative QCD corrections are shown in Fig.\ref{fg:d.sm} for the
${\cal SM}$ and Fig.\ref{fg:d.mssm} for the ${\cal MSSM}$.
They are large only in those region where strong destructive interferences are
present in the lowest order amplitude. This is rather dramatic in the
${\cal SM}$ where the top and $W$ loop are nearly cancelling each other
at a Higgs mass $m_H\sim 600$ Ge$\!$V.

\section{$\Phi\to gg$ [${\cal SM}$, ${\cal MSSM}$]}
%        =========================================
The gluonic decays $\Phi\to gg$ of the Higgs bosons in the ${\cal SM}$
and the ${\cal MSSM}$ are mediated in lowest order by loops of colored
particles with quarks providing the leading contributions. The lowest order
decay width is given by \cite{agaga,hgg}
\begin{eqnarray}
\Gamma(\Phi\rightarrow gg) = \frac{G_F \alpha_s^2}{36\sqrt{2}\pi^3} m_\Phi^3
\left| \sum_Q g_Q^\Phi A_Q^\Phi (\tau_Q) \right|^2
\end{eqnarray}
with the amplitudes
\begin{equation}
A_Q^H(\tau) = \frac{3}{2} \tau [1 +(1-\tau)f(\tau)] \hspace{3cm}
A_Q^A(\tau) = \frac{3}{2} \tau f(\tau)
\end{equation}
and $g_Q^\Phi$ denoting the corresponding couplings of Table \ref{tb:coup}.
The scaling variable $\tau_Q$ and the functions $f(\tau)$ are defined in the
previous section. Heavy quarks yield the dominant contribution to the decay
width, so that we restrict ourselves to the contributions of the top and
bottom quark in the following. In the ${\cal MSSM}$ the top quark part is
suppressed for large tg$\beta$, whereas the bottom one is enhanced in this
case. In the visible mass ranges the branching ratio of the gluonic decay
mode amounts to $\lsim 10^{-1}$.

The two-loop QCD corrections to the decay width can be parametrized by
\begin{equation}
\Gamma(\Phi\rightarrow gg(g),~gq\bar q) = \Gamma_{LO}(\Phi\rightarrow gg)
\left[ 1 + E_\Phi (\tau_Q) \frac{\alpha_s}{\pi} \right]
\end{equation}
The evaluation of the coefficients $E_\Phi (\tau_Q)$ requires the computation
of five-dimensional Feynman integrals for the virtual corrections, which have
been reduced analytically to one-dimensional ones containing trilogarithms
in the integrand. Ultraviolet, infrared and collinear singularities are
regularized in $n=4-2\epsilon$ dimensions. As in the photonic decay mode
the pseudoscalar $\gamma_5$ coupling is defined in the scheme of 't
Hooft-Veltman and Breitenlohner-Maison \cite{gamma5}. The counter terms are
fixed by defining the quark masses on-shell
and the strong coupling $\alpha_s$ in the $\overline{MS}$ scheme with five
active light flavors, i.e. the heavy top quark is decoupled. To obtain the
full QCD corrections the one-loop real corrections $\Phi\to ggg,~gq\bar q$
have to be added with phase space integration performed in $n$ dimensions.
Adding them to the virtual corrections infrared and collinear singularities
are cancelled resulting in finite corrections \cite{theo.h,hgg.qcd,hgg.qcd2}:
\begin{eqnarray}
E_H(\tau) = \frac{95}{4} - \frac{7}{6} N_F
+ \frac{33-2N_F}{6}\ \log \frac{\mu^2}{m_H^2} + \Delta E_H \nonumber \\
E_A(\tau) = \frac{97}{4} - \frac{7}{6} N_F
+ \frac{33-2N_F}{6}\ \log \frac{\mu^2}{m_A^2} + \Delta E_A
\end{eqnarray}
\begin{figure}[hbt]

\vspace*{-0.6cm}
\hspace*{0.90cm}
\begin{turn}{90}%
\epsfxsize=8.4cm \epsfbox{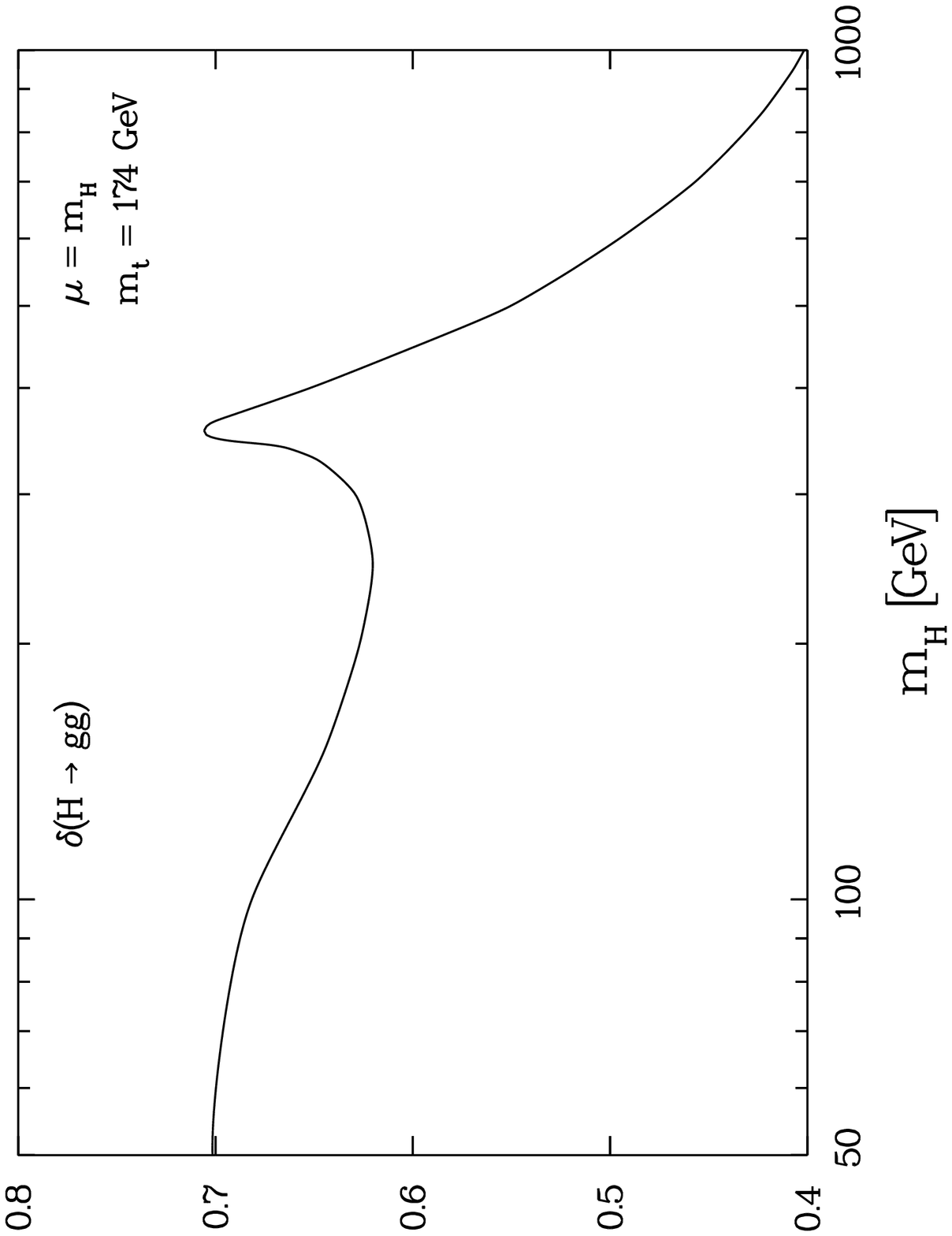}
\end{turn}
\vspace*{-0.3cm}

\fcaption{\label{fg:d.hgg} The relative QCD corrections to the gluonic decay
width of the ${\cal SM}$ Higgs boson.}
\vspace*{-0.2cm}

\end{figure}
\begin{figure}[hbt]

\vspace*{-0.45cm}
\hspace*{-1.30cm}
\begin{turn}{90}%
\epsfxsize=6.5cm \epsfbox{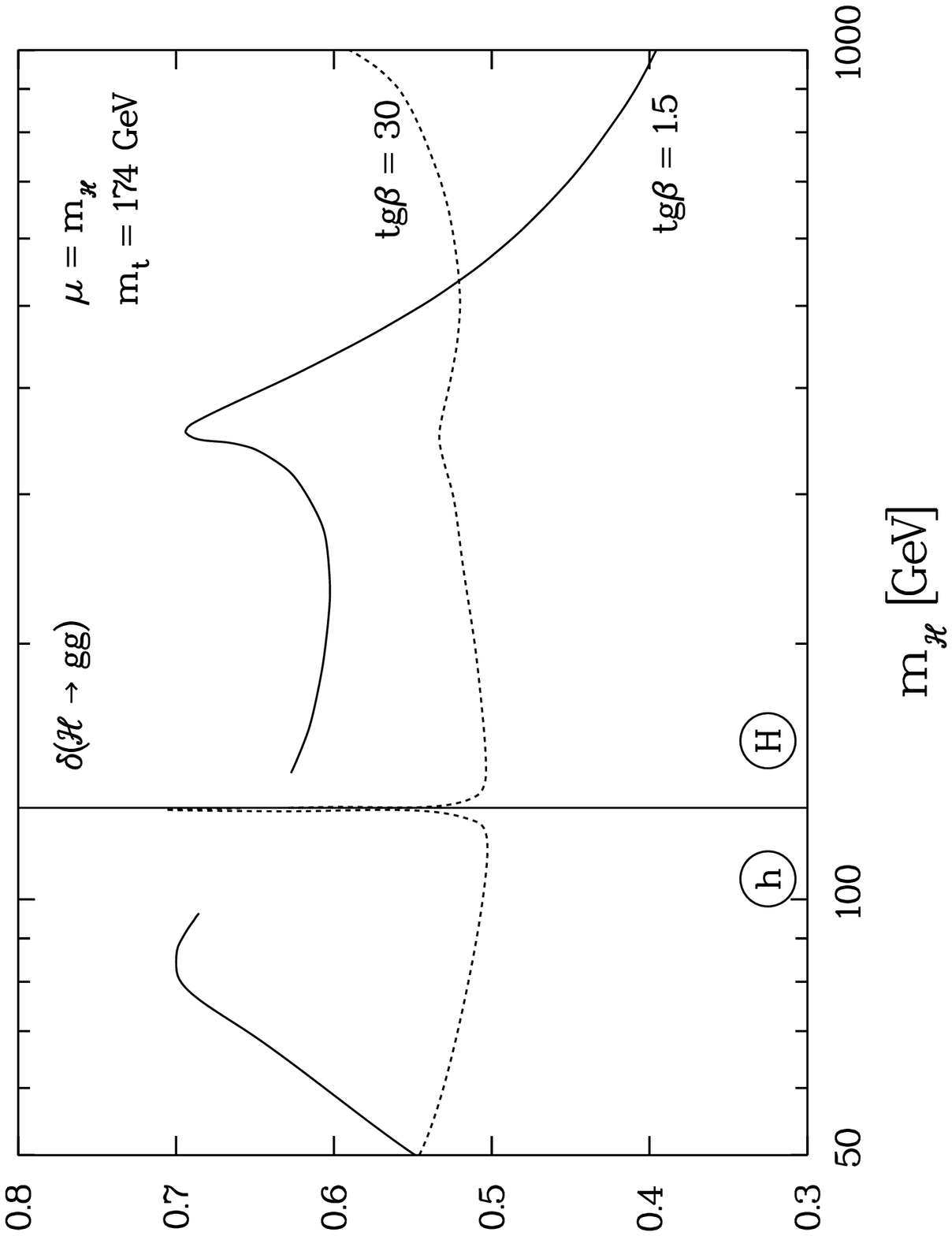}
\end{turn}
\vspace*{-0.8cm}

\vspace*{-5.73cm}
\hspace*{6.40cm}
\begin{turn}{90}%
\epsfxsize=6.5cm \epsfbox{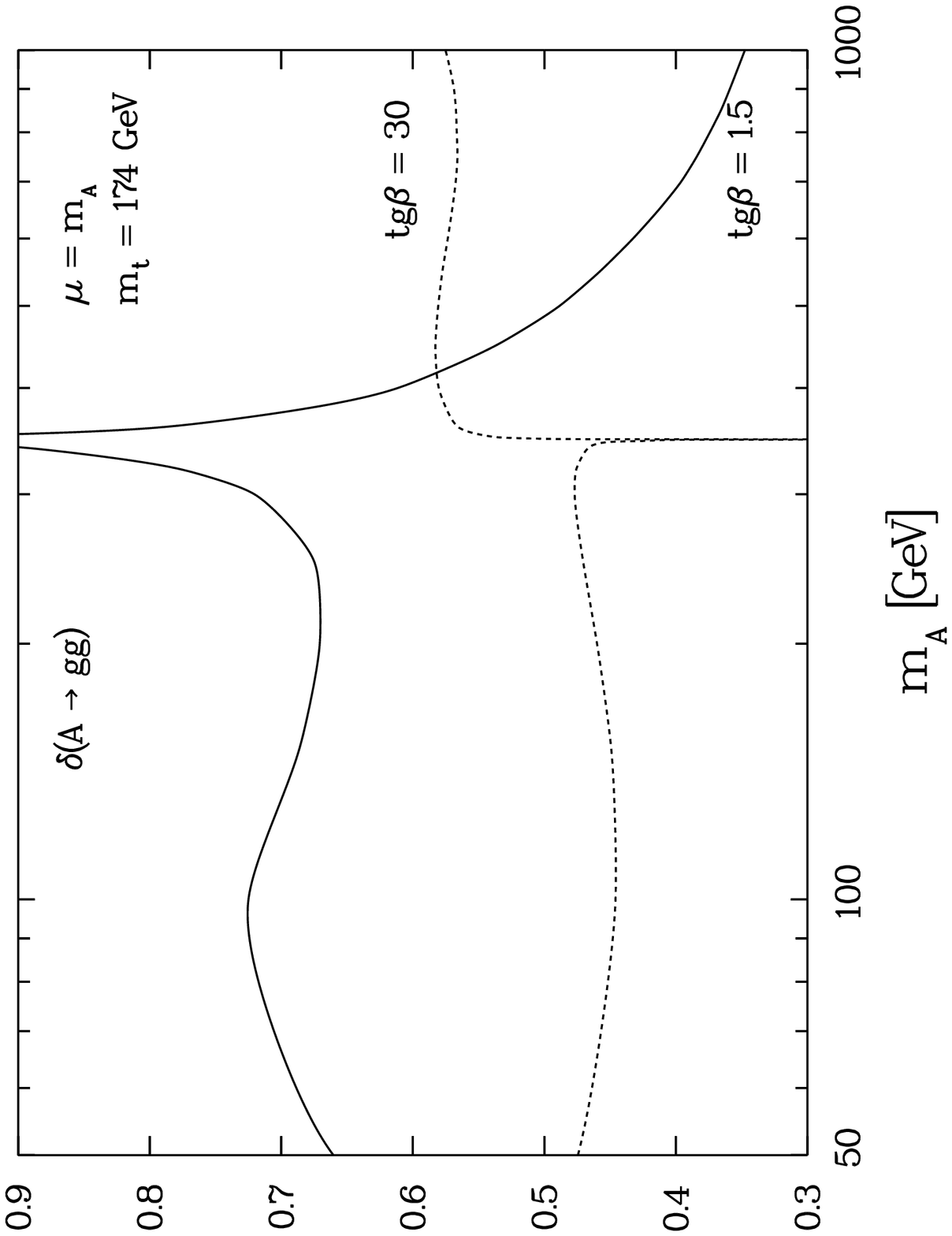}
\end{turn}
\vspace*{-0.3cm}

\fcaption{\label{fg:d.phigg} The relative QCD corrections to the gluonic decay
widths of the ${\cal MSSM}$ Higgs bosons for two different values of
tg$\beta$.}
\vspace*{-0.5cm}

\end{figure}
In the limit of heavy quark masses compared to the Higgs masses the
contributions $\Delta E_\Phi$ vanish; this can also be derived from
the low-energy theorems for scalar and pseudoscalar Higgs particles. The
QCD corrections to the pseudoscalar decay $A\to gg$ develop a Coulomb
singularity at threshold [$m_A=2m_Q$], so that the perturbative analysis is
not valid in a small margin around the threshold.
The QCD corrections amount to about 50 -- 70\% and
are shown in Fig.\ref{fg:d.hgg} for the ${\cal SM}$ Higgs boson and
Fig.\ref{fg:d.phigg} for the ${\cal MSSM}$ Higgs bosons. Hence they provide
an important contribution to the theoretical prediction of the gluonic
decay rates.


\section{References}
\begin{thebibliography}{99}
\vskip-1.5cm
\bibitem{higgs} P. W. Higgs, Phys. Rev. Lett. {\bf 12} (1964) 132 and Phys.
Rev. {\bf 145} (1966) 1156;
F. Englert and R. Brout, Phys. Rev. Lett. {\bf 13} (1964) 321;
G. S. Guralnik,	C. R. Hagen and	T. W. Kibble, Phys. Rev. Lett. {\bf 13}
(1964) 585.

\bibitem{janot} P. Janot, Talk delivered at {\it First General Meeting of
the LEP2 Workshop}, CERN, Geneva, Switzerland, 2--3 February 1995.

\bibitem{higbou} N. Cabibbo, L. Maiani, G. Parisi and R. Petronzio, Nucl. Phys.
{\bf B158} (1979) 295;
M. Chanowitz, M. Furman and I. Hinchliffe, Phys. Lett. {\bf B78} (1978) 285;
R. A. Flores and M. Sher, Phys. Rev. {\bf D27} (1983) 1679;
M. Sher, Phys. Rep. {\bf 179} (1989) 273;
M. Lindner, Z. Phys. {\bf C31} (1986)
295; \\
A.~Hasenfratz, K.~Jansen, C.~Lang, T.~Neuhaus and H.~Yoneyama, Phys. Lett.
{\bf B199} (1987) 531;
J.~Kuti, L.~Liu and Y. Shen, Phys. Rev. Lett. {\bf 61} (1988) 678;
M. L\"uscher and P. Weisz, Nucl. Phys. {\bf B318} (1989) 705.

\bibitem{cdf} F. Abe et al., CDF--Collaboration, Report
FERMILAB--PUB--95/022--E.

\bibitem{vacstab} M. Sher, Phys. Lett. {\bf B317} (1993) 159;
addendum {\bf B331} (1994) 448;
G. Altarelli and G. Isidori, Phys. Lett. {\bf B337} (1994) 141;
J. Casas, J. Espinosa and M. Quiros, Phys. Lett. {\bf B342} (1995) 171.

\bibitem{vacmeta}
J. Espinosa and M. Quiros, Report CERN-TH/95-18 and DESY 95-039.

\bibitem{eehgg}
Proceedings of the Workshop ``$e^+e^-$ Collisions at 500 Ge$\!$V: The
Physics Potential", DESY Report 92--123A; August 1992, P. Zerwas, ed.

\bibitem{hgagalhc} D. Froidevaux, Z. Kunszt and J. Stirling [conv.] et al. in
the Proceedings of the Large Hadron Collider Workshop, Aachen 1990,
CERN Report 90--10, Vol.~II, p. 427--603; see also the Rapporteurs talks
by G.~Altarelli and D.~Denegri in the same proceedings, Vol. I.

\bibitem{kunzwir}
Z.~Kunszt and F.~Zwirner, Nucl. Phys. {\bf B385} (1992) 3;
V.~Barger, M.~Berger, S.~Stange and R.~Phillips, Phys. Rev. {\bf D45} (1992)
4128;
H.~Baer, M.~Bisset, C.~Kao and X.~Tata, Phys. Rev. {\bf D46} (1992) 1067;
J.~Gunion and L.~Orr, Phys. Rev. {\bf D46} (1992) 2052.

\bibitem{gamfus} H. Haber and J. Gunion, 1990 DPF Summer Study on High Energy
Physics, SCIPP--90--22 and Phys. Rev. {\bf D48} (1993) 5109;
I. F. Ginzburg, Novosibirsk Preprint TF--28--182 (1990);
D.~L.~Borden, D.~A.~Bauer and D.~O.~Caldwell, Phys. Rev. {\bf D48} (1993)
4018;
M.~Kr\"amer, J.~K\"uhn, M. L.~Stong and P.~M.~Zerwas, Z. Phys. {\bf C64} (1994)
21;
D. Borden, V.A. Khoze, W.J. Stirling and J. Ohnemus, Phys. Rev. {\bf D50}
(1994) 4499.

\bibitem{mssmrad}
J. Gunion and A. Turski, Phys. Rev. {\bf D39} (1989) 2701 and {\bf D40} 2333;
M. Berger, Phys. Rev. {\bf D41} (1990) 225;
Y. Okada, M. Yamaguchi and T. Yanagida, Prog. Theor. Phys. {\bf 85} (1991) 1;
H. Haber and R. Hempfling, Phys. Rev. Lett. {\bf 66} (1991) 1815;
J. Ellis, G. Ridolfi and F. Zwirner, Phys. Lett. {\bf B257} (1991) 83;
R. Barbieri, F. Caravaglios and M. Frigeni, Phys. Lett. {\bf B258} (1991)
167;
A. Yamada, Phys. Lett. {\bf B263} (1991) 233;
J. R. Espinosa and M. Quiros, Phys. Lett. {\bf B266} (1991) 389;
A. Brignole, J. Ellis, G. Ridolfi and F. Zwirner, Phys. Lett. {\bf B271} (1991)
123;
P. H. Chankowski, S. Pokorski and J. Rosiek, Phys. Lett. {\bf B274} (1992)
191;
M. Drees and M. M. Nojiri, Phys. Rev. {\bf D45} (1992) 2482;
R. Hempfling and A. Hoang, Phys. Lett. {\bf B331} (1994) 99;
J. Espinosa and M. Quiros, Report CERN--TH--7334--94.

\bibitem{lepmssm}
ALEPH Collaboration, D.Buskulic et al., Phys. Lett. {\bf B313} (1993) 312;
L3 Collaboration, O. Adriani et al., Z. Phys. {\bf C57} (1993) 355;
OPAL Collaboration, R. Akers et al., Z. Phys. {\bf C64} (1994) 1;
DELPHI Collaboration, P. Abreu et al., Report CERN--PPE/94--218.

\bibitem{theorem1} J. Ellis, M.K. Gaillard and D.V. Nanopoulos, Nucl. Phys.
{\bf B106} (1976) 292.

\bibitem{theorem2} A. I. Va\u\i nste\u\i n, M. B. Voloshin, V. I. Sakharov and
M. A. Shifman, Sov. J. Nucl. Phys. {\bf 30} (1979) 711.

\bibitem{theorem} A.I. Va\u\i nshte\u\i n, V.I. Zakharov, and M.A. Shifman,
Usp.\ Fiz.\ Nauk {\bf131}, 537 (1980) [Sov.\ Phys.\ Usp.\ {\bf23}, 429
(1980)];
L.B. Okun, {\it Leptons and Quarks}, (North-Holland, Amsterdam, 1982) p.~229;
M.B. Voloshin, Yad.\ Fiz.\ {\bf44}, 738 (1986)
[Sov.\ J. Nucl.\ Phys.\ {\bf44}, 478 (1986)];
M.A. Shifman, Usp.\ Fiz.\ Nauk {\bf157}, 561 (1989)
[Sov.\ Phys.\ Usp.\ {\bf32}, 289 (1989)];
J.F. Gunion, H.E. Haber, G. Kane, and S. Dawson,
{\it The Higgs Hunter's Guide} (Addison-Wesley, Redwood City, 1990) p.~40.

\bibitem{theo.h} M. Spira, A. Djouadi, D. Graudenz and P.M.~Zerwas, Report
DESY--94--123.

\bibitem{theo.h2} B.A. Kniehl and M. Spira, Nucl. Phys. {\bf B432} (1994) 39.

\bibitem{delta.u} B.A. Kniehl, Phys. Rev. {\bf D50} (1994) 3314;
A. Djouadi and P. Gambino, Phys. Rev. {\bf D51} (1995) 218.

\bibitem{hbb.2} A. Kwiatkowski and M. Steinhauser, Phys. Lett. {\bf B338}
(1994) 66; Erratum {\it ibid.} {\bf B342} (1995) 455.

\bibitem{hbb.qcd}
E. Braaten and J.P. Leveille, Phys. Rev. {\bf D22} (1980) 715;
N. Sakai, Phys. Rev. {\bf D22} (1980) 2220;
T. Inami and T. Kubota, Nucl. Phys. {\bf B179} (1981) 171;
M. Drees and K. Hikasa, Phys. Lett. {\bf B240} (1990) 455.

\bibitem{drho2} A. Djouadi and C. Verzegnassi, Phys. Lett. {\bf B195} (1987)
265;
A. Djouadi, Nuovo Cim. {\bf 100A} (1988) 357;
B.A. Kniehl, Nucl. Phys. {\bf B347} (1990) 86.

\bibitem{pizz}
A. Djouadi and P. Gambino, Phys. Rev. {\bf D49} (1994) 3499, 4705.

\bibitem{drho1} D.A. Ross and M. Veltman, Nucl. Phys. {\bf B95} (1975) 135;
M. Veltman, Nucl. Phys. {\bf B123} (1977) 89.

\bibitem{hzz} B.A. Kniehl and M. Spira, Report DESY--95--008.

\bibitem{agaga} J.F. Gunion, G. Gamberini and S.F. Novaes, Phys. Rev.
{\bf D38} (1988) 3481.

\bibitem{gamma5} G. 't Hooft and M. Veltman, Nucl. Phys. {\bf B44} (1972)
189;
P. Breitenlohner and D. Maison, Commun. Math. Phys. {\bf 52} (1977) 11.

\bibitem{hgaga.ho} A.~Djouadi, M. Spira and P.~M.~Zerwas, Phys. Lett.
{\bf B311} (1993) 255.

\bibitem{hgaga.ho2}
H.~Zheng and D.~Wu, Phys.~Rev.~{\bf D42} (1990) 3760;
A.  Djouadi, M.  Spira, J.  van der Bij and P.  Zerwas, Phys.  Lett. {\bf B257}
(1991) 187;
S.~Dawson and R.~P.~Kauffman, Phys. Rev. {\bf D47} (1993) 1264;
K.~Melnikov and O.~Yakovlev, Phys. Lett. {\bf B312} (1993) 179;
M. Inoue, R. Najima, T. Oka and J. Saito, Mod. Phys. Lett. {\bf A9} (1994)
1189.

\bibitem{hgg.qcd}
A. Djouadi, M. Spira and P.M. Zerwas, Phys. Lett. {\bf B264} (1991) 440.

\bibitem{hgg.qcd1} S.~Dawson, Nucl. Phys. {\bf B359} (1991) 283.

\bibitem{abj} S.L. Adler, Phys. Rev. {\bf 177} (1969) 2416;
J. Bell and R. Jackiw, Nuovo Cim. {\bf 60A} (1969) 47.

\bibitem{adlbar} S.~L.~Adler and W.~A.~Bardeen, Phys.~Rev.~{\bf 182} (1969)
1517;
R. Jackiw, Lectures on Current Algebra and its applications, Princeton
University Press, 1972, New Jersey.

\bibitem{suther} D.G.Sutherland, Nucl. Phys. {\bf B2} (1967) 433.

\bibitem{thresh} K. Melnikov, M. Spira and O. Yakovlev, Z. Phys. {\bf C64}
(1994) 401.

\bibitem{hgg} H. Georgi, S. Glashow, M. Machacek and D. Nanopoulos,
Phys. Rev. Lett. {\bf 40} (1978) 692.

\bibitem{hgg.qcd2}
T. Inami, T. Kubota and Y. Okada, Z. Phys. {\bf C18} (1983) 69.

\end{thebibliography}
\end{document}